\documentclass[a4paper]{article}
\usepackage{graphicx}
\usepackage{dcolumn}
\usepackage{eucal}
\usepackage{amsfonts}
\usepackage{latexsym}
\usepackage{epsfig}
\usepackage{color}

\usepackage{amsfonts,amssymb,eucal}
\usepackage{amsthm}
 \usepackage{amsmath}
\usepackage{amscd}
 \usepackage{latexsym} 
\begin{document}
\title{Separation of unistochastic matrices from the double stochastic ones. Recovery of a $3\times3$ unitary matrix  from experimental data. }
\author{Petre Di\c t\u a$^{1}$\footnote {dita@zeus.theory.nipne.ro}\\
 Institute of Physics and Nuclear Engineering,\\
P.O. Box MG6, Bucharest, Romania}
\maketitle
\begin{abstract}\noindent

The aim of the paper is to provide a constructive method for recovering a unitary matrix from experimental data. Since there is a natural immersion of unitary matrices within the set of double stochastic ones, the problem to solve is to find  necessary and sufficient criteria that separate the two sets. A complete solution is provided for the 3-dimensional case, accompanied by a $\chi^2$ test necessary for the reconstruction of a unitary matrix from error affected data.

\end{abstract}
{\em keywords:} separation of unistochastic matrices from double stochastic ones, CP violation, recovery of unitary matrices from experimental data
\section{Introduction}

An $n\times n$ matrix $M$ is said to be  double stochastic if its elements satisfy the relations 
\begin{eqnarray} m_{ij}\ge 0,\quad \quad\sum_{i=1}^n\,m_{ij}=1,\quad\quad \sum_{j=1}^n m_{ij}=1\label{ds}\end{eqnarray}
All such  matrices form a convex set called the Birkhoff's polytope \cite{ Bir}. The unistochastic matrices are a subset of the double stochastic ones defined by
\begin{eqnarray} m_{ij}= |U_{ij}|^2\label{ds1}\end{eqnarray}
where $U$ is a unitary matrix, i.e. satisfies the relation
\begin{eqnarray}
U\,U^*=U^*\,U= I_n
\end{eqnarray}
where $*$ denotes the adjoint, and $I_n$ is the $n$-dimensional unit matrix.

It is well known that for $n\ge 3$ there are double stochastic matrices that are not unistochastic \cite{MO}, and  from a mathematical point of view there are a few  interesting problems that deserve to be solved:

$\imath)$ Given a double stochastic matrix, $M=(m_{ij})$, $i,j=1,\cdots,n$, what are the necessary and sufficient conditions for $M$ to be unistochastic.

$\imath\imath)$ Supposing  $M$ unistochastic, to what extent the matrix $U$ is determined by $M$, i.e. how many solutions one could get.

$\imath\imath\imath)$ If  $M$ is unistochastic, how one  can reconstruct the unitary $U$ from the given data.

The problem $\imath)$ was completely solved only for the 3-dimensional case \cite{YP}. A  characterization of the subset of double stochastic matrices that come from unistochastic ones, Eq.(\ref{ds1}), for $n=3$, was given in \cite{YP}, \cite{N}. For $n\ge 4$ only partial results are known, see e.g. \cite{MN}-\cite{BEKTZ}. As a matter of fact the theoretical physicists working on this problem were not aware of the embedding of the unitary matrices into the double stochastic ones, so in general they were not interested in obtaining necessary and sufficient conditions for the separation of the two sets. In this respect see  \cite{AMM}, Introduction,  where it is said: {\em We are not concerned here with the {\bf consistency} problem, which amounts to obtaining necessary and sufficient conditions on the set of numbers $|U_{jk}|$ for this set to represent the moduli of a unitary matrix}. In this paper we will provide such necessary and sufficient conditions  on the independent parameters entering the unitary matrix, when they are expressed in terms of the entries of a double stochastic matrix. We mention also that the Theorem 1 in  the paper \cite{YP},  which provides necessary and sufficient conditions in the $n=3$ case, is only an existence theorem, and it does not provide a  constructive method for recovering a unitary matrix from a double stochastic one.

Concerning  problem $\imath\imath)$ it was shown that the generic situation for   $n\ge 4$ is the existence of  a continuum of solutions, see \cite{A}-\cite{AMM} and \cite{D1}-\cite{BEKTZ}. In particular complex Hadamard matrices, i.e. unitary matrices with equal moduli, $|U_{ij}|= 1/\sqrt{n}$, have been found  that depend on arbitrary phases, see \cite{P}-\cite{TZ}. 

The solution of the mathematical problem $\imath\imath\imath)$ for $n=3$ can  be easily obtained, as we show in the following, but its applications in high energy physics put some challenges since the measured numbers are affected by errors. Usually one starts with a theoretical model formalized by a unitary matrix, and the problem is to recover the unitary matrix from measurements of its  moduli, or of some  ``angular looking objects'', $U_{i j}\, U_{l k}\, \overline{U}_{i k}\, \overline{U}_{l j} $, where bar means complex conjugation. When $|U_{ij}|^2$ are measured one gets a
  numerical matrix as 
\begin{eqnarray} V=(V_{ij}^2)\label{data}\end{eqnarray}
where every $V_{ij}$ is affected by errors, so we do not know how far from $|U_{ij}|$ are the measured values $V_{ij}$, and by  consequence the double stochasticity relations (\ref{ds}) are only approximately satisfied, if any.
More generally, that means that we have to test the full compatibility between the data and the unistochastic property. Second, by supposing  the data are  compatible to the existence of a unitary matrix we need a reliable  algorithm for an  explicit recovery  of  a unitary matrix from data, because some parameters entering the unitary matrix may have a physical significance, and  we want to know them, an example being the CP-violating phase from the  Cabibbo-Kobayashi-Maskawa (CKM) matrix. 

Indeed the recovery of a unitary matrix from experimental data is a central problem in the electro-weak interactions \cite{Ed} where the (assumed) unitary CKM matrix \cite{KM} plays a fundamental r\^ole.
Hence the recovery problem of a unitary matrix from experimental data is not an academic problem, it has many practical consequences. Let remind that there are two big Collaborations, BaBar in US, and Belle in Japan whose efforts are to measure as exact as possible the b-quark related entries of the CKM matrix. In Europe, at CERN, there is under construction the LHC machine, one of its  main aims  being a better understanding of the so called B-physics. Thus a reliable algorithm for   recovery of unitary matrices, able to obtain the independent parameters of the CKM matrix, could have consequences on both the design of future experiments, as well as on the design of future high energy machines, including neutrino factories.

The main goal of the paper is to provide a reliable algorithm for the reconstruction of a unitary matrix from experimental data, when these ones are compatible
with the theoretical model. We mention that the nowadays algorithm for reconstruction of a  $3\times 3$ unitary matrix from experimental data  does not make use of the double stochasticity relations (\ref{ds}), and the phenomenological model used to describe the data and to reconstruct a unitary matrix from them is mainly based on the use of   a {\em single} orthogonality relation, expressed as a triangle in the complex plane \cite{BaB}-\cite{Fl}, although  the proof of  Theorem 1 from paper \cite{YP} explicitly stresses the necessity of using at least {\em two} orthogonality relations.

   The paper is organized as follows. In sect.2, starting from the properties of double stochastic matrices and the embedding relation (\ref{ds1}) of unitary matrices within that set, we describe the gauge group of unitary matrices, i.e. the group of the most elementary transformations whose action on the unitary matrices does not change the unitarity property, or the physical content. In sect.3 we provide a parameterization of unitary matrices that is essential in devising a reconstruction algorithm in terms of physically relevant quantities.  In sect.4 we define two   phenomenological models, unitarity condition method, and unitarity triangles method, by using the embedding (\ref{ds1}) and the double stochasticity properties (\ref{ds}). We find  the necessary and sufficient conditions the data have to satisfy in each model in  order to be consistent with the unitarity properties, and give the reconstruction algorithm of unitary matrices from double stochastic matrices. In sect.5 we   show that both the approaches are completely equivalent, then and only then, when they are formulated in terms of {\em four} independent moduli; a consequence will be that the second model has to  use at least {\em two} orthogonality relations.  In sect.6 we describe the reconstruction algorithm of unitary matrices from experimental data that are compatible with the double stochasticity property. With this aim in view we define  $\chi^2$-tests that allow the recovery of unitary matrices from error affected data, including a method for doing statistics on moduli of unitary matrices. The paper ends by Conclusion.

~~~~~~~~~~~~~~~~~~~~~~~~~~~~~~~~~~~~~~~~~~~~~~~~
\section{Unitary matrices and their gauge subgroup}
~~~~~~~~~~~~~~~~~~~~~~~~~~~~~~~~~~~~~~~~~~~~~~~~

It is well known that an $n\times n$ unitary matrix depends on $n^2$ parameters,
\cite{M}, \cite{D3}, that are usually taken as $n(n-1)/2$ angles and $n(n+1)/2$ phases, each set taking values within $[0,\pi/2]$, and, respectively, $[0,2\,\pi)$.  Eq.(\ref{ds1}) tell us that, given a definite unitary matrix, all the unitary matrices obtained by multiplying  it at left and/or at right by  diagonal phase matrices, $D=diagonal\, (e^{i\,\phi_{1}}, \cdots, \,e^{i\,\phi_n} )$, with arbitrary real $\phi_i,\,\, i=1,\cdots,n$, generate a single  double stochastic matrix. That means that we can simplify a little bit the form of a unitary matrix since the values of $2 n -1$  phases are at our disposal, and a common choice for them is $0$ and/or $\pi$. In high energy physics this property is known as phase invariance \cite{CJ}-\cite{Br}. Thus we can take the entries from the first row and the first column as nonnegative quantities, such that the number of independent parameters entering a unitary matrix gets $ n^2-(2n-1)=(n-1)^2$, and it is equal with the number of independent parameters entering a double stochastic matrix. Hence we could say that the  embedding (\ref{ds1}) suggests   that the ``natural'' coordinates to parametrize a unitary matrix could be the moduli of its entries. Unfortunately that can be done only for $3\times 3$ matrices, but even in this case there are supplementary relations that have to be fulfilled by the moduli in order to get from them a unitary matrix.

Besides these transformations there are another transformations: multiplication at left and/or right by permutation matrices. Permutation matrices are  matrices whose elements on each row and each column are zero, but one that equals unity.  They interchange rows, and, respectively, columns between themselves. Both diagonal phase matrices and permutation matrices are subgroups of  unitary matrices. If $D$ denotes a diagonal phase matrix and $P$ a permutation matrix then
\[D\,\,D^* =P\,\,P^* = I_n\]
 Other equivalent unitary matrices can be obtained by taking the complex conjugate matrix, and/or the transpose of the original one. If we denote the transpose operator by $T$, and by $C$ the complex conjugation, these both transforms form a subgroup because
\[T^2 =C^2 = Identity\]
Thus the product group 
\[K=D \times P \times T \times C\]
is the gauge invariance subgroup of unitary matrices, and, in the following,  we work  only with the coset defined by
\begin{eqnarray}X  \cong U(n)/K\label{coset}\end{eqnarray}
where $U(n)$ denotes the $n$-dimensional unitary group. The $C$ invariance has an important consequence: the range of all independent phases entering $U(n)$ is  $[0,\pi]$. The multiplicity of solutions appearing in the recovery problem of a unitary matrix from a double stochastic one will be given modulo the above simplest transformations of unitary matrices.
~~~~~~~~~~~~~~~~~~~~~~~~~~~~~~~~~~~~~~~~~~~~~~~~~~~~~
\section{Parametrization of unitary matrices}
~~~~~~~~~~~~~~~~~~~~~~~~~~~~~~~~~~~~~~~~~~~~~~~~~~~~~
 
For devising a recovery algorithm of unitary matrices from double stochastic ones, or  from experimental data, we need an explicit parametrization of them. That will lead us easily to the necessary and sufficient conditions a double stochastic matrix has to satisfy for being also  unistochastic, even if a complete explicit solution is found only for $n=3$. These conditions will lead to   separation criteria between the double stochastic and unistochastic matrices.

There are essentially two types of parameterizations: first the classical result by Murnagham, \cite{M}, that states that the matrices from the   unitary group $ U(n)$ are products of a diagonal phase matrix containing $n$ phases, and $n(n-1)/2$ matrices whose main building block has the  form
\begin{eqnarray}
U=\left(\begin{array}{cc}
\cos\theta&-\sin\theta\, e^{-i\,\varphi}\\
\sin\theta\, e^{i\,\varphi}&\cos\theta\end{array}\right)
\end{eqnarray}
i.e. a parametrization in terms of $n(n-1)/2$ angles $\theta_i$, and $n(n+1)/2$ phases $\varphi_i$. The usual parametrization of the CKM matrix \cite{CK} used in high energy physics is of this type. A second parametrization, also through factorization, is that given  in paper \cite{D4}. The idea behind such a parametrization comes from the following sequence
\begin{eqnarray}\begin{split}
U(n)\cong{}& \frac{U(n)}{U(n-1)}\times\frac{U(n-1)}{U(n-2)}\times\dots\times\frac{U(2)}{U(1)}\times U(1)\\
\cong\,&\, S^{2n-1}\times S^{2n-3}\times\cdots\times S^3\times  S^1~\end{split}\label{fl}\end{eqnarray}
sequence that shows that each factor can be parameterized by an arbitrary point on the corresponding complex sphere, i.e. by a single complex $(n-k)$-dimensional unit vector. Such a parametrization could be appealing for high energy physicists since it shows that the information brought by each generation of quarks, or leptons is contained in a single complex unit vector. The first parametrization of CKM matrix, \cite{KM}, is of this form. The explicit realization of this parametrization, that is the main result in \cite{D4}, is given by

\newtheorem{Le} {Theorem}
\begin{Le}
 Any  element $U_n\in U(n)$ can be factored into an ordered product of $n$ matrices of the following form
\begin{eqnarray}{ U_n=B_n^0\cdots B_{n-1}^1\dots B_1^{n-1}\label{mat}}\end{eqnarray}
 where 

\begin{eqnarray}{ B_{n-k}^k=\left(\begin{array}{cc}
I_k&0\\
0&B_{n-k}
\end{array} \right)}\nonumber\end{eqnarray}
 and $B_{n-k}\in U(n-k)$ are special 
unitary matrices,  each one  generated by a single complex  $(n-k)$-dimensional unit vector, $b_{n-k}\in S^{2(n-k)-1}$. For example $B_1=e^{i\varphi}$, where $\varphi$ is an arbitrary phase.

If $y_m\in S^{2m-1},\,\,\, m=1,\cdots,n$, is parameterized by
\begin{eqnarray}{y_m=(e^{i\varphi_1}\cos\,\theta_1,e^{i\varphi_2}\sin\,\theta_1\cos\,\theta_2,\dots,e^{i\varphi_m}\sin\,\theta_1\dots \sin\,\theta_{m-1})^t\nonumber}\end{eqnarray}
where $t$ means transpose, then  the $m$ columns of $B_m$ are given by

\begin{eqnarray}
v_1=y_m=\left(\begin{array}{l}
e^{i\varphi_1}\cos\,\theta_1 \\
e^{i\varphi_2}\sin\,\theta_1\cos\,\theta_2 \\
\cdot\\
\cdot\\
\cdot\\
e^{i\varphi_m}\sin\,\theta_1\dots \sin\,\theta_{m-1}
\end{array} \right)\nonumber \end{eqnarray}
 and
\begin{eqnarray}{
v_{k+1}=\frac{d}{d\,\theta_k }\,\,v_1(\theta _1=\dots
=\theta _{k-1}=\pi/2),\qquad k=1,\cdots,m-1}, \nonumber \end{eqnarray}
where in the above formula one calculates first the derivative and afterwords
the restriction to $\pi/2$. \end{Le}

In what follows we need an explicit parametrization of  $B_{n-k}$, $k=0,\dots,n-1$, and by taking into account that we work with representatives from the X   coset, Eq.(\ref{coset}), we choose the corresponding $n$ generating vectors as follows
 
\begin{eqnarray}{\begin{array}{l}
y_n=(\cos\,a_1,\,\sin\,a_1\,\cos\,a_2,\cdots,\sin\,a_1\cdots \sin\,a_{n-1})^t\\
\\
y_{n-1}=(-\cos\,b_1,e^{i\beta_1}\sin\,b_1\,\cos\,b_2,\cdots,e^{i\beta_{n-2}}\sin\,b_1\cdots \sin\,b_{n-2})^t\\
....................................................................................................\\
y_2=(-\cos\,z_1,e^{i\psi_1}\sin\,z_1 )^t\\
\\
y_1=-1
\end{array}\label{Ve}}
\end{eqnarray}
With this choice the first row of $U_n$ has the form
\begin{eqnarray}u_{11}=\cos\, a_1,\quad u_{12}=\cos\, b_1\,\sin\,a_1,\quad\cdots\quad u_{1n}=\sin\, a_1\,\dots\, \sin\,z_1\label{row}\end{eqnarray}
and the first column is given by $y_n$.  In the following we will assume that $U_n$, Eq.(\ref{mat}), has its first column given by $y_n$, and its first row given by (\ref{row}), and this is our standard form in the rest of the paper for a unitary matrix. A similar parametrization was obtained recently, see \cite{J1}; for the case $n=3$ see also the papers \cite{CM} and \cite{MS}.

\section{Phenomenological models}

By ``phenomenological model'' we will understand in the following a relationship between the entries of a double stochastic matrix and the entries of a unitary matrix, the main goal being the finding of the necessary and sufficient conditions that separate the two sets. Depending on the context the ``experimental data'' will denote either  the entries of a double stochastic matrix, or the numbers, affected by errors, measured in an experiment.
Usually the experimental data on the $3\times 3$ CKM matrix entries from the quark sector  are given in terms of moduli of the unitary matrix that define the theoretical model, $|U_{ij}|$, or angular looking objects $U_{\alpha j}\, U_{\beta k}\, \overline{U}_{\alpha k}\, \overline{U}_{\beta j} $, that are equivalent with the angles of the triangles generated by the orthogonality relations, where bar means complex conjugation. For the beginning we assume that the data have no errors, which is the current mathematical setting, i.e. the moduli are the entries of a double stochastic matrix, and we want to solve the problems $\imath)- \imath\imath\imath)$ from Introduction. For doing that we have at our disposal an explicit parametrization of unitary matrices, (\ref{mat})-(\ref{Ve}), and the  unitarity property 
\[
U\,U^*=U^*\,U= I_n
\]
 This concise form  is equivalent with $2 n$ relations
\begin{eqnarray}
\sum_{i=1}^{i=n}|U_{ji}|^2-1=0, \quad j=1,\cdots,n\nonumber \\
\sum_{i=1}^{i=n} |U_{ij}|^2-1=0, \quad j=1,\cdots,n \label{sto}
\end{eqnarray}
 showing that the numbers
$ m_{ij}= |U_{ij}|^2$
 define a double stochastic matrix, that implies that only $2n-1$ relations from the set (\ref{sto}) are independent,  and by $ n(n-1)$ orthogonality relations
\begin{eqnarray}
\sum_{i=1}^{i=n} U_{ji}\,U_{ki}^*=0, \quad j < k,\quad i=1,\cdots,n\nonumber \\
\sum_{i=1}^{i=n} U_{ij}\,U_{ik}^*=0, \quad j < k,\quad i=1,\cdots,n\label{sto2}\end{eqnarray}
that can be visualized as polygons in the complex plane. The last relations are the supplementary relations the numbers $U_{ij}$ have to satisfy in order that the corresponding matrix should be unitary. The number of relations (\ref{sto}) and (\ref{sto2}) is greater than $n^2$, but we have written all of them since they could be useful in  over-constraining the experimental data, that usually are affected by errors.

The relations (\ref{sto}) and, respectively,  (\ref{sto2}) can be used to define two different phenomenological models. The first model is given by the relations
(\ref{sto}) together with 
\begin{eqnarray} m_{ij} = |U_{ij}|^2, \quad i,j=1,\cdots,n-1\label{ph1}\end{eqnarray}
where $ m_{ij}$ are the entries of a double stochastic matrix, and $U_{ij}$ are the entries of a unitary matrix parametrized as in  Theorem 1. The last relation is equivalent with the following equations
\begin{eqnarray}\begin{split}
m_{11}={}&\cos^2a_1,\, m_{12}=\sin^2a_1\,\cos^2b_1,\cdots,  m_{1n}=\sin^2a_1\cdots \sin^2z_1,\,m_{21}=\cos^2a_2\,\sin^2a_1,\\
 m_{31}=\,&\cos^2a_3\,\sin^2a_1\,\sin^2a_2,\cdots,m_{n\,n-1}=\sin^2a_1\cdots \sin^2a_{n-1},\end{split}\label{eq1}\end{eqnarray}
\begin{eqnarray}\begin{split}
 m_{22}={}&\cos^2a_1\cos^2a_2\cos^2b_1+\cos^2b_2 \sin^2a_2\sin^2b_1+\\
&2 \cos a_1\cos a_2 \cos b_1 \cos b_2\sin a_2 \sin b_1\cos\beta_1,\end{split}
\label{eq2}\end{eqnarray}
\begin{eqnarray}
\begin{split}
 m_{32}={}&\cos^2a_1\cos^2a_3\cos^2b_1 \sin^2a_2 +\cos^2 a_2\cos^2 a_3\cos^2 b_2 \sin^2 b_1+ \\
&\sin^2a_3\sin^2b_1\sin^2b_2-2 \cos a_1\cos a_2\cos^2a_3\cos b_1 \cos b_2 \sin a_2\sin b_1\cos \beta_1+\\
&2 \cos a_1\cos a_3\cos b_1\sin a_2\sin a_3\sin b_1\sin b_2\cos \beta_2-\nonumber\\
&2\cos a_2\cos a_3\cos b_2\sin a_3\sin^2 b_1\sin b_2\cos(\beta_1-\beta_2),\,\,\,{\rm etc.}\end{split}\\\label{eq3} \end{eqnarray}
where we have written only the simplest equations. It is easily seen from the above equations that, since $m_{1i}$ and $m_{i1}$, $i=1,\cdots,n-1$, are entries of a double stochastic matrix, there is a unique solution for $\cos a_i\in (0,1),\,\,i=1,\cdots,n-1$, of the form
\begin{eqnarray}
\cos^2a_1=m_{11},\quad\cos^2 a_k = \frac{m_{k1}}{1-\sum_{i=1}^{k-1} m_{i1}}\, ,\,\,\, k=2,\cdots,n-1 \label{q1}\end{eqnarray}
and similarly for $\cos b_1, \cos c_1,\cdots,\cos z_1$.  Hence the number of angles that have to be found is $(n-1)(n-2)/2 -(2n-3)=(n-2)(n-3)/2$. In that way the number of equations of the form (\ref{eq2})-(\ref{eq3})  we have to solve is only $(n-2)^2$. We substitute the forms for $\cos a_i,\,\sin a_i$, and those similar to, from the first column and the first row, in Eqs.(\ref{eq2})-(\ref{eq3}), such that we get $(n-2)^2$ equations that depend on $(n-1)^2$ moduli $m_{ij},\,i,j=1,\dots,n-1$. We do now a relabeling of the angles, $b_i\rightarrow b_{i-1},\,\,i=2,\cdots,n-2$, $c_2\rightarrow b_{n-1},\cdots$, etc., and similarly for the phases. With this notation the necessary and sufficient conditions for a double stochastic matrix to be also unistochastic are given by the relations
\begin{eqnarray} 0\le \cos b_i \le 1,\,\, i=1,\cdots, \frac{(n-2)(n-3)}{2}\label{comp1}\\
-1\le \cos \beta_i \le 1,\,\, i=1,\cdots, \frac{(n-1)(n-2)}{2}\label{comp2}
\end{eqnarray}
where $\cos b_i$ and $\cos\beta_i$ are the solutions in terms of $m_{ij}$ of the $(n-2)^2$ equations  of the form (\ref{eq2})-(\ref{eq3}). The above relations are in the same time the separation criteria between the double stochastic and unistochastic matrices, and their fulfillment is equivalent with the existence of at least one unitary matrix compatible with the moduli $m_{ij}$. To check them we have to solve analytically or numerically  the $(n-2)^2$ equations (\ref{eq2})-(\ref{eq3}). Numerically this can be done when $m_{ij}$ are the elements of a double stochastic matrix, but for the real case of experimental data with errors  we need an analytic solution to be used in a $\chi^2$-test, and until now this was found only for the case $n=3$. In conclusion for $3\times 3$ data coming from an exact double stochastic matrix we have only one constraint, namely,  $-1\le \cos\beta_1\le1$. If the data come from an experiment we have to check also the compatibility of the entries from the first row and first column with the relations (\ref{q1}), i.e. to see if the conditions $0\le \cos a_i\le 1$ are satisfied.

Taking into account the relation (\ref{ph1}), $m_{ij}= |U_{ij}|^2 $ , that shows how the unitary matrices are embedded into the double stochastic ones, 
the parameterization of  unitary matrices by their moduli seems to be  very appealing in this case, 
although it is not a natural one in the general case. A natural parameterization would be
one whose parameters are free, i.e. there are no supplementary constraints
upon them, as Eqs.(\ref{comp1})-(\ref{comp2}), to enforce unitarity. 

The problem we rose in \cite{D1} was to what extent the knowledge of the
moduli, $m_{ij}=|U_{ij}|$, of an $n\times n$ unitary matrix $U$ determines
$U$. If we identify the
parameters to the moduli, they will be lying within the simple domain
$$D=(0,1)\times\dots\times (0,1)\equiv (0,1)^{(n-1)^2}$$
where the above notation means that the number of factors entering the
topological product is $(n-1)^2$. We excluded only the extremities of each
  interval, i.e. the points  $0$ and $1$ that is a zero measure set within
${U}(n)$ and has no relevance to the problem of recovery of a unitary matrix from a double stochastic one.

  Nothing remains but to check if the new parameterization is
one-to-one. A solution to the last problem is the following: start with a
one-to-one parameterization of ${U}(n)$, as that given in the preceding section,  and then change the coordinates taking as
new coordinates the moduli of the $(n-2)^2$  entries; these ones are obtained by deleting the first and the last row, respectively, the first and the last column. The moduli of the first row and the first column are in one-to-one correspondence with the parameters entering the unitary matrix, see e.g. Eqs.(\ref{q1}), and the moduli entering the last row and the last column are uniquely determined by the double stochasticity property.  Afterwords use the implicit function theorem to find
the points where the new parameterization fails to be one-to-one. The
corresponding variety upon which the application is not a bijective one is
given by setting to zero the Jacobian of the transformation, i.e.
\begin{eqnarray}
J=\frac{\partial(m_{22},\dots,m_{2\, n-1},\dots,m_{n-1\,n-1})}{\partial({b_1,\dots,b_{(n-2)(n-3)/2},\beta_1,\dots,\beta_{(n-1)(n-2)/2})}}=0\end{eqnarray}
One gets that, generically, for $n\ge 4$ the unitary group $U(n)$ cannot be fully parameterized by the moduli of its entries, \cite{A}-\cite{TZ}, i.e. for a given set of moduli there exists a continuum of solutions, the  simplest example being  the case of complex Hadamard matrices \cite{P}-\cite{TZ}. The maximum dimension of the above variety is $(n-2)^2-1=(n-3)(n-1)$. For $n=3$, $J \ne 0$, and only in this case the parametrization of a unitary matrix through the moduli could be one-to-one. If the moduli are outside of the above variety an upper bound for the multiplicity is $2^{\frac{n(n-3)}{2}}$, bound that is saturated for $n=3$, when there is essentially only one complex matrix, if we take into account the gauge invariance of unitary matrices.

 In the case of exact double stochastic matrices, as we showed before, only $(n-2)^2$ moduli enter the game since the angles entering the first column and the first row are uniquely determined. To have a flavor of the problem we consider more in detail the  case $n=4$, when there are   four equations, two of them being Eqs.(\ref{eq2})-(\ref{eq3}), and the last two are 
\begin{eqnarray}
m_{23}=\cos^2 b_1 \cos^2 b_2 \cos^2 c_1\sin^2 a_2+\cos^2 a_1 \cos^2 a_2 \cos^2 c_1\sin^2 b_1+\nonumber\\
\sin^2 a_2 \sin^2 b_2 \sin^2 c_1-2 \cos a_1  \cos a_2 \cos b_1 \cos b_2 \cos c_1\sin a_2 \sin b_1 \cos \beta_1+\nonumber\\
2 \cos b_1  \cos b_2 \cos c_1 \sin^2 a_2 \sin b_2  \sin c_1 \cos \gamma_1-~~~~~~~~~~~~~~~~~~~~~~~~~~~~~~~~~~~~~\nonumber\\
2 \cos a_1  \cos a_2 \cos c_1 \sin a_2 \sin b_1 \sin b_2  \sin c_1 \cos(\beta_1+ \gamma_1)~~~~~~~~~~~~~~~~~~~~~~~~~\label{eq5}
\end{eqnarray}
\begin{eqnarray}
m_{33}=\cos^2a_2 \cos^2a_3 \cos^2b_1 \cos^2b_2 \cos^2 c_1+\cos^2 a_1 \cos^2 a_3  \cos^2 c_1 \sin^2 a_2 \sin^2 b_1\nonumber\\
+\cos^2b_1\cos^2c_1\sin^2 a_3 \sin^2 b_2+\cos^2b_2\sin^2a_3\sin^2c_1 +\cos^2a_2 \cos^2a_3 \sin^2 b_2 \sin^2 c_1\nonumber\\
+2 \cos a_1 \cos a_2 \cos^2 a_3 \cos b_1 \cos b_2  \cos^2 c_1 \sin a_2 \sin b_1\cos \beta_1~~~~~~~~~~~~~~~~~\nonumber\\
-2 \cos a_2 \cos a_3 \cos^2 b_1 \cos b_2  \cos^2 c_1\sin a_3 \sin b_2\cos( \beta_1-\beta_2)~~~~~~~~~~~~~~~~\nonumber\\
-2\cos a_1 \cos a_3 \cos b_1  \cos^2 c_1\sin a_2\sin a_3\sin b_1\sin b_2\cos\beta_2~~~~~~~~~~~~~~~~~~\nonumber\\
+2\cos a_2 \cos a_3 \cos b_1 \cos^2 b_2\cos c_1\sin a_3 \sin c_1\cos(\beta_1-\beta_2-\gamma_1)~~~~~~~~~~\nonumber\\
+2 \cos a_1 \cos a_3 \cos b_2 \cos c_1\sin a_2 \sin a_3\sin b_1\sin c_1\cos(\beta_2+\gamma_1)~~~~~~~~~~~\nonumber\\
+2 \cos^2 a_2 \cos^2 a_3\cos b_1 \cos b_2 \cos c_1\sin b_2 \sin c_1\cos\gamma_1~~~~~~~~~~~~~~~~~~~~~~~~\nonumber\\
-2 \cos b_1 \cos b_2 \cos c_1\sin^2a_3\sin b_2 \sin c_1\cos\gamma_1~~~~~~~~~~~~~~~~~~~~~~~~~~~~~~~~~\nonumber\\
+2 \cos a_1 \cos a_2\cos^2 a_3\cos c_1 \sin a_2 \sin b_1 \sin b_2 \sin c_1\cos(\beta_1+\gamma_1)~~~~~~~~~\nonumber\\
-2\cos a_2 \cos a_3 \cos b_1 \cos c_1\sin a_3 \sin^2b_2 \sin c_1\cos(\beta_1-\beta_2+\gamma_1)~~~~~~~~~~\nonumber\\
+2  \cos a_2\cos a_3\cos b_2 \sin a_3  \sin b_2 \sin^2 c_1\cos(\beta_1-\beta_2)~~~~~~~~~~~~~~~~~~~~~~~~\label{eq6}\end{eqnarray}

From Eqs.(\ref{eq1}) we get
\begin{eqnarray}
\cos a_1&=&\sqrt{m_{11}},~~~~~\quad ~~~~~~~~~~\sin a_1=\sqrt{1-m_{11}}\nonumber\\
\cos a_2&=&\sqrt{\frac{m_{21}}{1-m_{11}}},~~~~\quad~~~~ \sin a_2=\sqrt{\frac{1-m_{11}-m_{21}}{1-m_{11}}}\nonumber\\
\cos a_3&=&\sqrt{\frac{m_{31}}{1-m_{11}-m_{21}}},\quad \sin a_3=\sqrt{\frac{1-m_{11}-m_{21}-m_{31}}{1-m_{11}-m_{21}}}\label{eq7}\\
\cos b_1&=&\sqrt{\frac{m_{12}}{1-m_{11}}},~~~\quad ~~~~~\sin b_1=\sqrt{\frac{1-m_{11}-m_{12}}{1-m_{11}}}\nonumber\\
\cos c_1&=&\sqrt{\frac{m_{13}}{1-m_{11}-m_{12}}},\quad \sin c_1=\sqrt{\frac{1-m_{11}-m_{12}-m_{13}}{1-m_{11}-m_{12}}}\nonumber
\end{eqnarray}

By substituting relations (\ref{eq7}) into equations (\ref{eq2})-(\ref{eq3}) and  (\ref{eq5})-(\ref{eq6}) we obtain four equations that depend on  nine moduli $m_{ij},\,\,i,j=1,2,3$, and on the non-relabeled parameters $b_2,\,\beta_1,\,\beta_2,\,\gamma_1$
\begin{eqnarray} f_{22}-m_{22}&=&0, \quad f_{23}-m_{23}=0,\nonumber\\
f_{32}-m_{32}&=&0, \quad f_{33}-m_{33}=0\label{eq8}\end{eqnarray}
In general the rank of the Jacobian matrix
\begin{eqnarray}
J=\frac{\partial(f_{22},f_{23},f_{32},f_{33})}{\partial(b_2,\beta_1,\beta_2,\gamma_1)}\label{jac1}\end{eqnarray}
 will be less than $(n-2)^2$ since we know that there are particular solutions that depend on an arbitrary phase. If we look at the Jacobian (\ref{jac1}) as a function on the Birkoff's polytope, i.e. depending on  $(n-1)^2=9$ independent moduli $m_{ij}$ it could be possible to find a domain where  rank($J)=4$, i.e. in this case we have only one solution. Then the 
compatibility relations between the  double stochastic matrix  entries $(m_{ij})_{i,j=1}^3$, and the unitarity property, or in other words, the separation criteria between the two sets, are four, and they have the form 
\begin{eqnarray}
0\le \cos b_2\le1,\quad -1\le \cos \beta_1\le 1,\quad -1\le \cos \beta_2\le 1,\quad -1\le \cos \gamma_1\le 1\label{eq9}\end{eqnarray}

The above relations are  the necessary and sufficient conditions the moduli of a $4\times 4$ double stochastic matrix have to satisfy in order to exist a unitary matrix whose moduli coincide with $m_{ij}$, and their intersection gives the maximal domain within the  $m_{ij}$ simplex that is compatible to the existence of unitary matrices.

In case when for a given numerical matrix, rank($J)< 4$,
there is no one-to-one correspondence between the entries $m_{ij}$ of a double stochastic matrix and the independents parameters entering a unitary matrix, i.e. there is at least one solution that depends on an arbitrary parameter, phase or angle. 

  The checking of criteria (\ref{eq9}) requires explicit analytic solutions for \[ \cos b_2,\,\cos \beta_1,\,\cos \beta_2,\,\cos\gamma_1\] in terms of $m_{ij}$, and when these ones are not numbers, solving the Eqs.(\ref{eq8}), is not a simple problem even with the symbolic calculation software packages nowadays available. The only results in this direction are those obtained in \cite{AMM}, however they have to be used with caution since the authors assumed that no matter how the numbers entering a double stochastic matrix are, the Eqs.(\ref{eq8}) have a physical solution.

By taking into account the above considerations the following result holds
\setcounter{Le}{1}
\begin{Le}
Suppose we have   a generalized spherical coordinate system on the unitary
group ${U}(n)$, and let  $U\in U(n)$ be a given matrix parameterized as in  Theorem 1, through $n(n-1)/2$ angles, each one taking
values in $[0,\pi/2]$, and $(n-1)(n-2)/2$ phases taking values in $[0,\pi]$, and let $M=(m_{ij})$ be a $n\times n$ double stochastic matrix, whose entries are supposed to come from a unistochastic matrix $U$ by the embedding 
\begin{eqnarray} m_{ij}={|U_{ij}|^2,\,\, i,j=1,\cdots,n-1\label{eq10}}\end{eqnarray}
From equations (\ref{eq1}) we get a unique solution for the angles entering the first column and the first row of $U$ as follows
\begin{eqnarray}\cos^2a_1&=& m_{11},\,\,\cos^2a_2=\frac{m_{21}}{1-m_{11}},\cdots,\cos^2a_{n-1}=\frac{m_{n-1\,1}}{1-m_{11}-\sum_{i=2}^{n-1} m_{i1}},\nonumber\\
\cos^2 b_1&=&\frac{m_{12}}{1-m_{11}},\cdots,\cos^2 z_1=\frac{m_{1\,n-1}}{1-m_{11}-\sum_{i=2}^{n-1}m_{ 1\,i}}\label{eq11}\end{eqnarray}
We substitute them in Eqs.(\ref{eq10}) obtaining a set of $(n-2)^2$ equations that, after relabeling of the angles and phases, is of the form
\begin{eqnarray}
f_{ij}(b_1,\dots,b_{\frac{(n-2)(n-3)}{2}}, \beta_1, \cdots, \beta_{\frac{(n-1)(n-2)}{2}})=m_{ij},\,\, i,j=2,\cdots, n-1\label{eq12}\end{eqnarray}
The solutions of the above equations are compatible with the existence of a unitary matrix, if and only if, all the angles and phases satisfy the unitarity constraints (\ref{comp1})-(\ref{comp2}).
For $n\ge 4$ the solutions of equations (\ref{eq12}) could depend on arbitrary
angles and/or phases on the
variety obtained by setting to zero the determinant of Jacobian matrix of the transformation (\ref{eq12})
\begin{eqnarray}
J={\partial(f_{22},\cdots,f_{2\,n-1},f_{n-1\,2}\cdots,f_{n-1\,n-1})\over\partial(b_1,\cdots,b_{\frac{(n-2)(n-3)}{2}},\beta_1,\cdots,\beta_{\frac{(n-1)(n-2)}{2}})}\label{jac} \end{eqnarray}

If $p$ is the rank of the Jacobian matrix (\ref{jac})
the  solution of (\ref{eq12}) depends upon  $(n-2)^2-p$
arbitrary parameters, angles and/or phases.
Outside this variety the number of discrete  solutions $N_s$ satisfies $1\leq
N_s\leq 2^{{n(n-3)\over 2}}$.
 \end{Le}

{\em Proof.} Since we use a spherical coordinate system the equations (\ref{eq12}) are trigonometric equations in our parameterization, as the example of case $n=4$ shows, Eqs.(\ref{eq2}),(\ref{eq3}),(\ref{eq5}),(\ref{eq6}), and consequently the multiplicity of the solutions may arise from the
two possible phase solutions
 for all values of sine or cosine functions that
satisfy Eqs.(\ref{eq12}) and the constraints (\ref{comp1})-(\ref{comp2}).  The number of independent phases is $(n-1)(n-2)/2$
and, since we do not make any distinction between $U$ and its complex conjugate $\bar{U}$,
 condition that halves the number of solutions, the above
bound for $N_s$ follows.

 For $n=3$ the Jacobian does not vanish and one gets $1\leq  N_s\leq 1$, and this bound
implies the existence of a complex unitary matrix, then and only then, when the values $m_{ij}$, coming from a double stochastic matrix satisfy the relation $-1\le\cos\beta_1 \le 1$. $\Box$
\vskip1mm
An example of a unitary matrix that cannot be recovered from its moduli is the following. If $P$ and $Q_i,\,i=1,\cdots,m$, are $m\times m$ and, respectively, $n\times n$ unitary matrices whose first rows and first columns are positive numbers and depend on $p$ and respectively $q_i$ arbitrary phases, then the following $m\times n$ array
\begin{eqnarray}
M=\left(\begin{array}{cc cc}
p_{11}\, Q_1&\cdot&\cdot&p_{1m}\,Q_m\\
\cdot&\cdot&\cdot&\cdot\\
\cdot&\cdot&\cdot&\cdot\\
p_{1m}\, Q_1&\cdot&\cdot&p_{mm}\,Q_m\end{array}\right)\end{eqnarray}
defines a unitary matrix that could depend on 
\begin{eqnarray}
p+q_1+(m-1)\sum_{i=2}^n\,q_i
\end{eqnarray}
arbitrary phases.

Indeed it is easily seen that we can multiply at left all the matrices $Q_2,\cdots,Q_m$ by diagonal phase matrices $D_j=Diagonal(1,e^{\imath\,\varphi_{1,j}},\cdots, ,e^{\imath\,\varphi_{n-1,j}}),\, j=2,\cdots,m$, by preserving the entries from the first row and column  of $M$ positive numbers, obtaining a set of unitary matrices that are all applied in the same double stochastic matrix. Thus their recovery from the moduli cannot be done because all the arbitrary phases $\varphi_{ij}$ disappear when one computes  their moduli, and by consequence they do not appear in equations as (\ref{eq2})-(\ref{eq3}). 

A second phenomenological model can be defined by starting from the orthogonality relations (\ref{sto2}), but since in this case the polygons angles enter the game its formulation for arbitrary $n$ is more difficult. Thus in the following we will discuss the case $n=3$ for  both the models, that has  applications in high energy physics. 
 
\subsection{Unitarity condition method}

In the $n=3$ case the unitary (CKM) matrix is parametrized by {\em four} independent parameters given by the so called mixing angles, $\theta_{12},\theta_{13},\theta_{23}$, and the $CP$-violating phase $\delta$. Hence in the following we will change the notation from section 3 to another notation that is more familiar to experimenters and phenomenologists. That means that in  equations (\ref{Ve}) defining the generating vectors we make the substitution: $(a_1\rightarrow\theta_{12},\,a_2 \rightarrow\theta_{23},\,b_1\rightarrow\theta_{13},\,\beta_1\rightarrow\delta)$; after that we make the notation
\[\cos\theta_{ij}=c_{ij},\,\sin\theta_{ij}=s_{ij},\,\,ij=12,13,23\]
 and by using  Theorem 1 we get the following form
\begin{eqnarray}
U=\left(\begin{array}{lcc}
c_{12}&c_{13}s_{12}&s_{12}s_{13}\\
c_{23}s_{12}&-c_{12}c_{13}c_{23}-e^{i\,\delta}s_{13}s_{23}&-c_{12}c_{23}s_{13}+e^{i\,\delta}c_{13}s_{23}\\
s_{13}s_{23}&c_{23}s_{13}e^{i\,\delta}-c_{12}c_{13}s_{23}&-c_{13}c_{23}e^{i\,\delta}- c_{12}s_{13}s_{23}\end{array}\right)\label{ckm}\end{eqnarray}
The theoretical model (\ref{ckm}) is supplemented by the experimental data supplied by experimenters. In the quark sector one measures two kinds of parameters: the moduli of the unitary matrix (\ref{ckm}), see \cite{Ed},  under the form of a positive entries matrix, written with the physicists notation
\begin{eqnarray}
V=\left(\begin{array}{ccc}
V_{ud}^2&V_{us}^2&V_{ub}^2\\*[1mm] 
V_{cd}^2&V_{cs}^2&V_{cb}^2\\*[1mm] 
V_{td}^2&V_{ts}^2&V_{tb}^2\\
\end{array}\right)\label{pos}
\end{eqnarray}
where $u,\,s,\,b$, etc. are names for quarks, and the angles of the so called standard unitarity triangle, \cite{BaB}, denoted by $\alpha/\phi_1,\,\beta/\phi_2,\gamma/\phi_3$, \cite{HF}. In this paper $V$ denotes either a double stochastic matrix, or a set of numbers affected by errors, when $V_{ij}$ are measured in  experiments.
 More generally the experimental data can be given  in terms of some functions $f_k(V_{ij}),\,k=1,\dots,N$, that depend on the $V$ entries, or the theoretical parameters $s_{ij}$ and $\delta$.

Similarly to the general case treated in the previous section, we define our phenomenological model as a relationship  between the theoretical object (\ref{ckm}) and the experimental data (\ref{pos}). It is given by 
 the double stochasticity relations (\ref{sto}), which now  take the form
\begin{eqnarray}
\sum_{i=d,s,b} V_{ji}^2-1=0, \quad j=u,c,t\nonumber \\
\sum_{i=u,c,t} V_{ij}^2-1=0, \quad j=d,s,b \label{sto1}
\end{eqnarray}
and by the embedding relation of a unitary matrix into the double stochastic set
\[V=|U|^2\]
that leads to the following relations
\begin{eqnarray}
V_{ud}^2&=&c^2_{12},\,\, V_{us}^2=s^2_{12}c^2_{13},\,\,V_{ub}^2=s^2_{12} s^2_{13}\nonumber \\
 V_{cd}^2&=&s^2_{12} c^2_{23},\,\,
 V_{td}^2=s^2_{12} s^2_{23},\nonumber\\
V_{cs}^2&=&c^2_{12}c^2_{13} c^2_{23}+s^2_{13} s^2_{23}+2 c_{12}c_{13}c_{23}s_{13}s_{23}\cos\delta,\nonumber\\
V_{cb}^2&=&c^2_{12} c^2_{23} s^2_{13}+c^2_{13} s^2_{23}-2 c_{12}c_{13}c_{23}s_{13}s_{23}\cos\delta\label{unitary},\\
V_{ts}^2&=&c^2_{23}s^2_{13}+c^2_{12}c^2_{13}s^2_{23}-2 c_{12}c_{13}c_{23}s_{13}s_{23}\cos\delta\nonumber,\\
V_{tb}^2&=& c^2_{13} c^2_{23}+c^2_{12}s^2_{13}s^2_{23} +2 c_{12}c_{13}c_{23}s_{13}s_{23}\cos\delta\nonumber
\end{eqnarray}
The above  phenomenological model was introduced in \cite{D5}.
  The relations (\ref{unitary}) depend only on $\cos\delta$ which has the consequence that we can restrict $\delta$ to the interval $[0,\,\pi]$, this property
being equivalent to the CKM matrix invariance under the complex conjugation, as it was shown in section 2.

Especially for physicists we want to make a few remarks. First we  stress that in any phenomenological analysis one 
works with {\em two} distinct objects: the first is the theoretical one,
that in our case coincides with the matrix $U$, Eq.(\ref{ckm}), which  is {\em assumed and built as a unitary matrix}; the second object is provided by the
 {\em experimental data}, $V=(V_{ij}^2)$, Eq.(\ref{pos}). The aim of any phenomenological analysis is twofold: a) {\em  checking  the   consistency of   data  with the theoretical model}, and, b)  {\em determination of parameters entering the theoretical model from the experimental data, if these ones are consistent with it}.  That is the reason for making  a clear distinction between  the theoretical quantities and the experimental ones, by using different symbols for denoting them. The second remark concerns the double stochasticity relations, Eqs.(\ref{sto1}), that are considered by (many) high energy physicists as testing the {\em unitarity}, statement which is wrong, since it is well known that for $n\ge 3$ there exist double stochastic matrices which are not unistochastic, \cite{MO}.
Checking the consistency of the data with the theoretical model means checking the consistency of relations (\ref{unitary}), i.e. we have to see if the solutions of Eqs. (\ref{unitary}) lead to physical values for the mixing parameters $\theta_{ij}$ and $\delta$; and only in this case  Eqs.(\ref{sto1}) {\em together} with Eqs.(\ref{unitary}) prove the unitarity property of the data.

Let us assume for a moment that the relations (\ref{sto1}) are exactly satisfied. Then it is an easy matter to find from the first five relations (\ref{unitary}) three independent ones which give  a unique solution for the $c_{ij},\, ij=12,13,23$. In other words, if the experimental numbers satisfy the relations
\[V_{ud}^2+V_{us}^2+V_{ub}^2=1\]
\[V_{ud}^2+V_{cd}^2+V_{td}^2=1\]
we get always a solution for $c_{ij}$ that is unique and depends on the three chosen independent parameters.
 Substituting this solution in the last equations one gets four equations for
 $\cos\delta$, that lead to a unique solution for it. But nobody guarantees us that the solution will satisfy the physical constraint
\begin{eqnarray}
 -1\leq\cos\delta\leq 1\label{unit1}
\end{eqnarray}
 The last relation gives   the necessary and sufficient condition 
the data have to satisfy in order  the $3\times 3$ matrix (\ref{pos}) comes from a unitary matrix, i.e. it is the consistency condition between the data and the theoretical model.

To better understand the above considerations and see the power of the found criterion (\ref{unit1}) and how it works, we will give a few numerical examples, and for that we will use  moduli entering the first two rows.  We make the following notation:  \[V_{ud}=a,\,\,V_{us}=b,\,V_{ub}=c,\,\, V_{cd}=d,\,\,  V_{cs}=e,\,\,{\rm and}\,\, V_{cb}=f\]
First we choose as independent moduli $a,b,d,\,\,{\rm and}\,\,e$ and with them  form the square root of a double stochastic matrix
\begin{eqnarray}
S_1=\left(
\begin{array}{ccc}
a&b&\sqrt{1-a^2-b^2}\\*[1mm]
d&e&\sqrt{1-d^2-e^2}\\*[1mm]
\sqrt{1-a^2-d^2}&\sqrt{1-b^2-e^2}&\sqrt{-1+a^2+b^2+d^2+e^2}
\end{array}\right)\label{toy1}
\end{eqnarray}
 i.e. $S_1^2$ is an exact doubly stochastic matrix, where the square is taken entry wise, by using the Hadamard product from linear algebra.  From the  relations (\ref{unitary}) we get the solution
\begin{eqnarray}c_{12}=V_{ud}=a,\,\, c_{13}=\frac{V_{us}}{\sqrt{1-V_{ud}^2}}=
\frac{b}{\sqrt{1-a^2}},\,\, c_{23}=\frac{V_{cd}}{\sqrt{1-V_{ud}^2}}=
\frac{d}{\sqrt{1-a^2}}\label{rel}\end{eqnarray}
\begin{eqnarray}
\cos\delta_1=\frac{-(1-a^2)^2(1-e^2)+(1-a^2)(b^2+d^2)-b^2 d^2(1+a^2))}{2 a b d \sqrt{1-a^2-b^2} \sqrt{1-a^2-d^2}}\label{cosd1}
\end{eqnarray}

In the second case we take  $b,\,\,c,\,\,d,\,\,{\rm and },\,\,f$ as independent moduli,  and get 
\begin{eqnarray}
S_2=\left(
\begin{array}{ccc}
\sqrt{1-b^2-c^2}&b&c\\*[1mm]
d &\sqrt{1-d^2-f^2}&f\\*[1mm]
\sqrt{b^2+c^2-d^2}&\sqrt{d^2+f^2-b^2}&\sqrt{1-c^2-f^2}
\end{array}\right)\label{toy2}
\end{eqnarray}
\begin{eqnarray}
c_{12}=\sqrt{1-b^2-c^2},\,\,c_{13}=\frac{b}{\sqrt{b^2+c^2}},\,\,c_{23}=\frac{d}{\sqrt{b^2+c^2}}\label{rel1}\end{eqnarray}

\begin{eqnarray}
\cos\delta_2=\frac{b^2(b^2+c^2)-d^2(b^2-c^2+c^2(b^2+c^2))-f^2(b^2+c^2)^2}{2  b c d \sqrt{1-b^2-c^2} \sqrt{b^2+c^2-d^2}}\label{cosd2}
\end{eqnarray}

If the data are the entries of the following double stochastic  matrix 
\begin{eqnarray}
V=\left(\begin{array}{ccc}
\frac{1}{3}&\frac{1}{2}&\frac{1}{6}\\*[2mm]
\frac{1}{4}&\frac{2}{5}&\frac{7}{20}\\*[2mm]
\frac{5}{12}&\frac{1}{10}&\frac{29}{60} \end{array}\right)\label{toy3}\end{eqnarray}
we get from the equations (\ref{rel})-(\ref{cosd1}), and (\ref{rel1})-(\ref{cosd2})
\begin{eqnarray}
c_{12}=\frac{1}{\sqrt{3}},\,\,c_{13}=\frac{\sqrt{3}}{2},\,\,c_{23}=\frac{\sqrt{6}}{4},\,\,
\cos\delta_1=\cos\delta_2=\frac{4\sqrt{15}}{25}\label{del1}\end{eqnarray}
and the results show  that the data are compatible to the existence of a unitary matrix. We remark that no matter how the independent moduli are chosen, $c_{ij}$ and $\cos\delta$ takes the same value, and this is a consequence of the fact that the properties of a double stochastic matrix do not change by multiplying it at left and/or right by permutation matrices. From a mathematical point of view the story ends here, because we can easily reconstruct the unitary matrix whose moduli are given in (\ref{toy3}), by using the results (\ref{del1}) in the unitary matrix (\ref{ckm}). We get

\begin{eqnarray}
U=\left(\begin{array}{ccc}
\frac{1}{\sqrt{3}}&\frac{1}{\sqrt{2}}&\frac{1}{\sqrt{6}}\\*[2mm]
\frac{1}{2}&-\frac{9}{20}\sqrt{\frac{3}{2}}-\frac{1}{20}\sqrt{\frac{77}{2}}\,i&
\frac{7}{20\sqrt{2}}+\frac{1}{20}\sqrt{\frac{231}{2}}\,i\\*[2mm]
\frac{1}{2}\sqrt{\frac{5}{3}}&-\frac{13}{20\sqrt{10}}+\frac{1}{20}\sqrt{\frac{231}{10}}\,i&-\frac{61}{20\sqrt{30}}-\frac{3}{20}\sqrt{\frac{77}{10}}\,i
 \end{array}\right)\label{unt}\end{eqnarray}

Hence the reconstruction algorithm of unitary matrices from the double stochastic ones is the following: {\em Start with a double stochastic matrix as {\rm (\ref{toy3})} and solve the system of equations  {\rm(\ref{unitary})}. If the numerical value for $\cos\delta$ satisfies the inequalities  {\rm(\ref{unit1})}, then with the values for $c_{ij}$ and  $\cos\delta$ go to  {\rm(\ref{ckm})} and find the corresponding unitary matrix}.
    
 For experimental data as those recommended in \cite{Ed} the situation changes. For example, by using the numbers: $a=0.9738\pm0.0005 ,\,b=0.22\pm 0.0026,\,c=0.00367\pm 0.00047,\,d=0.224\pm 0.012,\,e=0.996\pm 0.013,\,f=0.0423\pm0.0015 $, to  define two doubly stochastic matrices $S_1$ and $S_2$ one gets
\begin{eqnarray}\cos\delta_1^+&=& -0.03\,i,\quad \cos\delta_1^c=1.59,\quad \cos\delta_1^-=1.08 \\
\cos\delta_2^+&=&8.95\,i,\quad \cos\delta_2^c=5.985\,i,\quad\cos\delta_2^-=7.699\label{inc1}\end{eqnarray}
where the indexes $+,\,c,\,-$ denote the $\cos\delta$ values obtained from central values$+1 \sigma$, the central values, and, respectively,  central values$-1 \sigma$.
The above  results  show  that our criterion (\ref{unit1}) is very sensitive  to small variations of the parameters of the order of errors, and, on the other hand, one sees that the fulfillment of 
 unitarity for experimental data is not an easy problem.  We remark that $\cos\delta_1\ne \cos\delta_2$, although both the matrices (\ref{toy2}) and  (\ref{toy1}) lead to double stochastic matrices, but these ones are different because numerically, e.g. $e\ne \sqrt{1-d^2-f^2}$. Hence in the case of experimental data we have to take care and try to find how the necessary and sufficient conditions for the existence of a unitary matrix could be implemented. In this case also the relations (\ref{sto1}) are not exactly fulfilled. Consequently the numbers  $c_{ij}$ obtained  from the relations (\ref{rel}) and (\ref{rel1}) could be different,  depending on the independent parameters we use for their determination.
On the other hand the last four relations (\ref{unitary}) provide us formulas for $\cos\delta$ and these formulas have to give the same number when comparing theory with experiment, by supposing the data come from a unitary matrix. Their explicit form depends on the independent four parameters  we choose to parameterize the data. In fact there are 58 independent groups of four independent moduli that lead to 165 different expressions for  $\cos\delta$.  Depending on the explicit choice of the four independent parameters we get one, two, three or four  different expressions for $\cos\delta$.

Looking at equations  (\ref{rel})-(\ref{cosd1}) and  (\ref{rel1})-(\ref{cosd2}) we see that the expressions defining the mixing angles and phase $\delta$ are quite different. Thus if the data are compatible to the existence of a unitary matrix these  angles $c_{ij}$ and phases $\delta^{(i)}$ have to be equal, and these are  the most general necessary  conditions for unitarity; they  can be written as
\begin{eqnarray}0\le c_{ij}^{(m)}\le 1, \,\, c_{ij}^{(m)}=c_{ij}^{(n)},\,\,m, n=1,\cdots,58,\,\, \cos\delta^{(i)}=\cos\delta^{(j)},\,i,j=1,\cdots,165\nonumber\end{eqnarray}
The  above relations are also satisfied  by the double stochastic matrices, and the condition that separates the unitary matrices from the  double stochastic ones is given by the relation (\ref{unit1}), i.e. $-1\le \cos\delta^{(i)}\le 1$.

\subsection{Unitarity triangle method}

The second phenomenological model is defined by 
 the orthogonality relations of rows, and, respectively, columns of a unitary matrix, and by the double stochastic relations (\ref{sto1}).
Although there are six such relations, see Eqs.(\ref{sto2}), usually one considers only the orthogonality of the first and the third columns of $U$, relation that is written as
\begin{eqnarray}
U_{ud} U_{ub}^* + U_{cd} U_{cb}^* + U_{td} U_{tb}^*=0\label{ort}
\end{eqnarray}
The above equation can be visualized as a triangle in the complex plane.
Usually (\ref{ort})  is  scaled by dividing it 
  through the middle term such that the length of one side is 1. Taking into account that  our parametrization, Eq.(\ref{ckm}), of a unitary matrix has the entries of the first column and the first  row positive quantities, we  divide by the first term, $U_{ud} U_{ub}^*$, which is positive. In fact what matters are the angles of the triangle, and our choice has the advantage that the triangle generated by Eq.(\ref{ort}) has two angles which numerically  coincide with the phases of $U_{23}$ and   $U_{33}$; together with the  phases of $U_{22}$ and   $U_{32}$ they can be used for the determination of the unitary matrix $U$, since  all these angles are measurable quantities in experiments, see e.g. \cite{BaB}, or \cite{AKL}. 

The other sides have the lengths
\begin{eqnarray}
R_{db,c}^{(1)}&=&\left|\frac{U_{cd} U_{cb}^*}{U_{ud} U_{ub^*}}\right|=
\frac{d\,\sqrt{1-d^2-e^2}}{a\,\sqrt{1-a^2-b^2}}\label{tri1}\\
R_{db,t}^{(1)}&=&\left|\frac{U_{td} U_{tb}^*}{U_{ud} U_{ub}^*}\right|=\frac{\sqrt{1-a^2-d^2}\,\,\sqrt{a^2+b^2+d^2+e^2-1}}{a\,\sqrt{1-a^2-b^2}}\nonumber
\end{eqnarray}
where we have written on the right hand side the R-values in our choice of the four independent parameters by using  the  matrix  (\ref{toy1}). The physical condition takes the form
\begin{eqnarray}
|R_{db,c}^{(1)}-R_{db,t}^{(1)}|\le 1\le  R_{db,c}^{(1)}+R_{db,t}^{(1)} \label{tri2}\end{eqnarray}
that says that with the lengths $1,\,R_{db,c}$ and $R_{db,t}$ one can construct a triangle.
If we use the matrix (\ref{toy2}) we find
\begin{eqnarray}
R_{db,c}^{(2)}&=&\left|\frac{U_{cd} U_{cb}^*}{U_{ud} U_{ub}^*}\right|=
\frac{d\,f}{c\,\sqrt{1-b^2-c^2}}\label{tri3}\\
R_{db,t}^{(2)}&=&\left|\frac{U_{td} U_{tb}^*}{U_{ud} U_{ub}^*}\right|=\frac{\sqrt{1-c^2-f^2}\,\,\sqrt{b^2+c^2-d^2}}{c\,\sqrt{1-b^2-c^2}}\nonumber
\end{eqnarray}

We remark that in Eqs.(\ref{tri1})-(\ref{tri3}) the left side is the same, and only the right side differs, because the four independent moduli we use are different. Since there are 58 different groups of independent moduli there will be 58 different expressions  $R_{db,j}^{(i)},\,\, j=c, t, \,\,{\rm and}\,\,i=1,\cdots,58$.

If now we use the orthogonality between the first and the second columns, i.e.
\begin{eqnarray}
U_{ud} U_{us}^* + U_{cd} U_{cs}^* + U_{td} U_{ts}^*=0\label{ort1}
\end{eqnarray}
 one gets similarly 
\begin{eqnarray}
R_{ds,c}^{(1)}&=&\left|\frac{U_{cd} U_{cs}^*}{U_{ud} U_{us}^*}\right|=\frac{d\,e}{a\,b}\label{tri4}\\
R_{ds,t}^{(1)}&=&\left|\frac{U_{td} U_{ts}^*}{U_{ud} U_{us}^*}\right|=\frac{\sqrt{1-a^2-d^2}\,\,\sqrt{1-b^2-e^2}}{a\,b}\nonumber
\end{eqnarray}
and respectively
\begin{eqnarray}
R_{ds,c}^{(2)}&=&\left|\frac{U_{cd} U_{cs}^*}{U_{ud} U_{us}^*}\right|=\frac{d\,\sqrt{1-d^2-f^2}}{b\,\sqrt{1-b^2-c^2}}\label{tri5}\\
R_{ds,t}^{(2)}&=&\left|\frac{U_{td} U_{ts}^*}{U_{ud} U_{us}^*}\right|=\frac{\sqrt{b^2+c^2-d^2}\,\sqrt{d^2+f^2-b^2}}{b\,\sqrt{1-b^2-c^2}}\nonumber \end{eqnarray}
 If as in the preceding case we compute the expressions on the right hand side using the data (\ref{toy3}) we find 
\begin{eqnarray}
R_{db,c}^{(1)}=R_{db,c}^{(2)}=\frac{3}{2}\sqrt{\frac{7}{10}},\quad 
R_{db,t}^{(1)}=R_{db,t}^{(2)}=\frac{1}{2}\sqrt{\frac{29}{2}}
\end{eqnarray}
and respectively
\begin{eqnarray}
R_{ds,c}^{(1)}=R_{ds,c}^{(2)}=\sqrt{\frac{3}{5}},\quad
R_{ds,t}^{(1)}=R_{ds,t}^{(2)}=\frac{1}{2}\end{eqnarray}
 that both satisfy the inequalities of the form (\ref{tri2}).

With the central  values from  \cite{Ed} one gets
\begin{eqnarray}
\begin{array}{llll}
R_{db,c}^{(1)}=0.8\,i&R_{db,t}^{(1)} =0.71&R_{ds,c}^{(1)}= 1.04&R_{ds,t}^{(1)} =0.037\,i\\
R_{db,c}^{(2)} =2.58&R_{db,t}^{(2)} =11.72\, i&R_{ds,c}^{(2)}=1.016 &R_{ds,t}^{(2)} =0.012\,i \end{array}\label{inc2} \end{eqnarray}
result that sends the same signal of incompatibility as in the previous subsection, see Eqs.(\ref{inc1}).

From the above equations we can obtain the angles of the triangles generated by the relations (\ref{ort}) and, respectively, (\ref{ort1}). For each triangle we denote by  $\varphi_3 $ the angle of the  triangle associated to the vertex $(0,0)$, the other two,   $\varphi_1$ and $\varphi_2$, being associated respectively  to the vertexes $(\rho,\eta)$ and $(1,0)$, where $(\rho,\eta)$ are the  coordinates of the triangle apex. For the second triangle we make
the substitution
$\varphi_i\rightarrow \psi_i,\,i=1,2,3$, and find
\begin{eqnarray}
\cos\varphi_1=4\sqrt{\frac{7}{145}}\approx0.8,\,\,\cos\varphi_2=\frac{61}{10\sqrt{58}}\approx0.88,\,\,\cos\varphi_3=-\frac{1}{2}\sqrt{\frac{7}{10}}\approx-0.42\label{tri11}\end{eqnarray}
and respectively
\begin{eqnarray}
\cos\psi_1=-\frac{1}{4}\sqrt{\frac{3}{5}}\approx -0.19,\quad\cos\psi_2=\frac{13}{20}\approx0.65,\quad\cos\psi_3=\frac{9}{8}\sqrt{\frac{3}{5}}\approx0.87\label{tri12}\end{eqnarray}

Similarly to the preceding case the necessary  and sufficient conditions for unitarity are the constraints: all  $R^{(j)}\ge 0$, and
\begin{eqnarray}R_{db,c}^{(i)}=R_{db,c}^{(j)},\quad R_{db,t}^{(i)}=R_{db,t}^{(j)},\,\,\,i,j=1,\cdots,58\label{tri6}\\
 R_{ds,c}^{(i)}= R_{ds,c}^{(j)},\quad R_{ds,t}^{(i)}=R_{ds,t}^{(j)},\,\,\,i,j=1,\cdots,58\label{tri7}\end{eqnarray}
\begin{eqnarray}
|R_{db,c}^{(i)}-R_{db,c}^{(j)}|\le 1 \le R_{db,c}^{(i)}+R_{db,c}^{(j)} ,\,\,j=1,\cdots,58\label{tri8}\\
|R_{ds,t}^{(i)}-R_{ds,t}^{(j)}|\le 1 \le R_{ds,t}^{(i)}+R_{ds,t}^{(j)} \label{tri9},\,\,j=1,\cdots,58\end{eqnarray}
where in the last equations we have written only the conditions implied by two orthogonality relations, although for applications we must calculate the constraints for all the six orthogonality relations.

The first remark is that this approach, in the variant used by  physicists,  does not make use of the double stochasticity relations, Eqs.(\ref{sto1}), the physicists implicitly assuming that they are satisfied by the numbers obtained from experiments. With the above notation, $(\rho,\eta)$, for the apex of the triangle,  Eqs.(\ref{tri1}) are written under the form, see e.g. \cite{BaB}, \cite{BLO}-\cite{Fl}
\begin{eqnarray}
R_{db,c}^{(1)}&=&\left|\frac{U_{cd} U_{cb}^*}{U_{ud} U_{ub}}\right|=
\sqrt{\rho^2+\eta^2}\label{tri10}\\
R_{db,t}^{(1)}&=&\left|\frac{U_{td} U_{tb}^*}{U_{ud} U_{ub}}\right|=\sqrt{(1-\rho)^2+\eta^2}\nonumber
\end{eqnarray}

The second remark is that in contradistinction to what physicists believe, $\rho$ and $\eta$ have no special relationship with a  parametrization of the CKM matrix, in particular  that provided by Wolfenstein \cite{W}. Indeed on the right hand,  Eqs.(\ref{tri10}), are the lengths of two sides of the above defined triangle, those $\ne 1$. Physical meaning  have only the angles of that triangle which can be measured  \cite{BaB}, \cite{AKL}.  The third  remark is that
 in this approach there is no relationship between the $CP$-violating phase $\delta$ and the angles of the unitarity triangles. Because the phase is interesting from a physical point of view, the phenomenologists make the identification 
\begin{eqnarray}\delta=\varphi_3,\quad {\rm or} \,\,\delta\approx\varphi_3\label{del2}\end{eqnarray}
see e.g. \cite{BLO}-\cite{Fl}. 
Looking at the numerical values obtained  for $\cos\delta=\frac{4}{25}\sqrt{15}$, relation (\ref{del1}), and  for the angles of the unitarity triangles, (\ref{tri11})-(\ref{tri12}), computed by using an exact unitary matrix, we arrive at the conclusion that the claim (\ref{del2}) is definitely wrong. 

A simpler example is the following: take all $V_{ij}^2=1/3$. Then all the six triangles are equilateral and by consequence we have $\phi_1= \phi_2=\phi_3=60^{\circ} $, and from the first phenomenological model we get $\delta=90^{\circ}$.

Although this model in the form (\ref{tri10}) is currently used in many phenomenological analyzes, see e.g. \cite{JC} and \cite{Bo}, it cannot provide directly numbers for the parameters $c_{ij}$ and $\delta$ such that there is no (reliable) recovery algorithm for unitary matrices from double stochastic matrices. The positive thing is that, if properly used, this phenomenological model allows the determination of all the angles of all the six unitarity triangles, which are measurable quantities. Hence the real problem is to find a recovery algorithm for unitary matrices starting with measured values for all these angles. This problem was first raised by Aleksan {\em et al}, \cite{AKL}, and in the next section we will solve it.

\section{Equivalence of the two approaches }

 The relations (\ref{inc1}) and  (\ref{inc2}), as well as (\ref{del1}) and (\ref{tri11}), have shown that the unitarity sends the same signal of (in)consistency between the data and the theoretical model, although each one  in a specific way. This is natural since both the models are based on the unitarity property of matrices modeling the $CP-$ violation.  In the following we will prove that from a theoretical point of view the above phenomenological models  are only {\em partially} equivalent in the following sense: if we start with four independent angles we can reconstruct more than a unitary matrix, the multiplicity being equal to five,  {\em  and the solutions given by the two models are the same if and only if  four moduli take the same values}. This opens the possibility to define new phenomenological models, in terms of moduli and phases, by taking into account all the experimental data. In particular 
we  provide an expression for the phase $\delta$ in the second phenomenological model.

The starting point is the relation (\ref{tri1}) that we write in a complex form. For that we define the four independent angles that enter the CKM matrix (\ref{ckm}), namely $U_{ij}=|U_{ij}|e^{\imath\, \omega_{ij}}$, $i,j=2,3$. In our notation, (\ref{tri1}), $\omega_{23}$ is the angle located at $(0,0)$ between the positive x-axis and the complex vector $U_{cd}^* U_{cb}/U_{ud}^* U_{ub}$ oriented towards the apex $(\rho,\eta)$;  $\pi-\omega_{33}$ is the angle located at $(1,0)$ between the positive $x$-axis and the complex vector $U_{td}^* U_{tb}/U_{ud}^* U_{ub}$ oriented also to the same apex. With this notation the complex form of (\ref{tri1}) is written as 
\begin{eqnarray}
R_{db,c}\,(\cos \omega_{23} + i \sin \omega_{23}) =  \,\frac{U_{cd}^*\,U_{cb}}{U_{ud}^*\,U_{ub}}\label{l1}\\
R_{db,t}\,(-\cos \omega_{33} + i \sin \omega_{33}) =  \, \frac{U_{td}^*\,U_{tb}}{U_{ud}^*\,U_{ub}}\label{l2}\end{eqnarray}
where from we get
\begin{eqnarray}
\sin \omega_{23}& =&\frac{c_{13}\,c_{23}\,s_{23}\,\sin\delta}{c_{12}\,s_{13}\,  R_{db,c}}\label{l3}\\
\cos \omega_{23}&=&-\frac{c_{23}(c_{12}\,c_{23}\,s_{13}-c_{13}\,s_{23}\,\cos\delta)}{c_{12}\,s_{13}\, R_{db,c}}\label{l4}\end{eqnarray}
The above relations are equivalent with 
\begin{eqnarray}
\tan \omega_{23}=\frac{c_{13}\,s_{23}\,\sin\delta}{-c_{12}\,c_{23}\,s_{13}+c_{13}\,s_{23}\,\cos\delta}\label{l5}\end{eqnarray}
 The last formula depends only on theoretical parameters entering (\ref{ckm}), and  does not depend on the lengths of the unitarity triangle.
 If in (\ref{l5}) we 
 substitute values for $c_{ij}$ and $\cos \delta$  taken for example from the phenomenological model, 
Eqs.(\ref{rel})-(\ref{cosd1}), we get a formula for $\tan \omega_{23}$ in terms of moduli that are measurable quantities. 

 Conversely, from the relations (\ref{l3})-(\ref{l4}) we get a formula for $\cos \delta$. Indeed, from the identity $\sin^2 \omega_{23}  +\cos^2  \omega_{23}=1$
we find
\begin{eqnarray}
\cos \delta = \frac{c_{12}^2\,c_{23}^4\,s_{13}^2+ c_{13}^2\,c_{23}^2\, s_{23}^2-c_{12}^2\, s_{13}^2\,R_{db,c}^2 }{2\,c_{12}\,c_{13}\,c_{23}^2\,s_{13}\,s_{23}}
\label{l6}\end{eqnarray}
 If in it we substitute the formulas (\ref{rel}) and
 $R_{db,c}$ from Eqs.(\ref{tri1}) we find Eq.(\ref{cosd1}) for $\cos \delta$. If instead of  (\ref{tri1})
we use the corresponding form for $R_{db,c}$, that comes from the relation (\ref{tri3}), one gets Eq.(\ref{cosd2}), and so on. The above formula shows that if we want to use the lengths of the unitarity triangle to obtain $\cos\delta$ we have to provide values for $c_{ij}$ from an other source.

In the same way from the relation (\ref{l2}) one gets
\begin{eqnarray} \tan \omega_{33}=\frac{c_{13}\,c_{23}\,\sin \delta}{c_{12}\,s_{13}\,s_{23}+c_{13}\,c_{23}\cos \delta}\label{l7}\end{eqnarray}
and
\begin{eqnarray} \cos\delta =\frac{c_{12}^2\,s_{13}^2\, R_{db,t}^2 -c_{13}^2\,c_{23}^2\,s_{23}^2-c_{12}^2\,s_{13}^2\,s_{23}^4}{2\,c_{12}\,c_{13}\,c_{23}\,s_{13}\,s_{23}^3}\label{l8}\end{eqnarray}
The third angle is given by $\pi-\omega_{23}-\omega_{33}$. Of course from the  relations (\ref{l1})-(\ref{l2}) we find 
\begin{eqnarray}\frac{R_{db,t}}{R_{db,c}}(-\cos(\omega_{23}+\omega_{33}) + i\,\sin(\omega_{23}+\omega_{33}) )=\frac{U_{td}^*\,U_{tb}}{U_{cd}^*\,U_{cb}}\end{eqnarray}
and from it we can obtain a similar formula for $\tan(\omega_{23}+\omega_{33})$.

Similarly the complex form of the second triangle, Eqs.(\ref{tri4}), is

\begin{eqnarray}
R_{ds,c}\,(\cos \omega_{22} + i \sin \omega_{22}) =  \,\frac{U_{cd}^*\,U_{cs}}{U_{ud}^*\,U_{us}}\label{l9}\\
R_{ds,t}\,(-\cos \omega_{32} + i \sin \omega_{32}) =  \, \frac{U_{td}^*\,U_{ts}}{U_{ud}^*\,U_{us}}\label{l10}\end{eqnarray}
where from we get
\begin{eqnarray} \tan \omega_{22}=\frac{s_{13}\,s_{23}\,\sin\delta}{c_{12}\,c_{13}\,c_{23}+s_{13}\,s_{23}\,\cos\delta}\label{l11}\end{eqnarray}
\begin{eqnarray} \tan \omega_{32}=\frac{c_{23}\,s_{13}\,\sin\delta}{-c_{12}\,c_{13}\,s_{23}+c_{23}\,s_{13}\,\cos\delta}\label{l12}\end{eqnarray}

From the above calculations one sees that an orthogonality relation determines only two independent angles, and from them one cannot reconstruct the unitary matrix because each angle depends on the {\em four} independent parameters entering the generic form (\ref{ckm}) of a $3\times 3$ unitary matrix. 

From a mathematical point of view the angles $\omega_{ij},\,i,j=2,3$ are not very interesting, although  their existence was the essential ingredient for obtaining the necessary and sufficient conditions for the existence of a unitary matrix from the entries of a double stochastic matrix, see \cite{YP}. From a physical point of view they are very interesting because they are measurable quantities. Similarly to the preceding cases we have to use all the six orthogonality relations, although for a double stochastic matrix all the angles  $\omega_{ij},\,i,j=2,3$ have the same numerical values, irrespective of the orthogonality property we use. However irrespective what triangles we use we get  the same functions $\tan \omega_{22}, \,\tan \omega_{23},\,\tan \omega_{32},  $ and $\tan \omega_{33}$. If we use  the orthogonality of the second and the third columns, respectively of the second and the third rows, we get the angles of the corresponding triangles as linear functions of $\omega_{22},\,\omega_{23},\,\omega_{32},\,\omega_{33}$, see \cite{AKL}.

In the following we give the necessary and sufficient conditions for recovery of a unitary matrix when we know the angles  $\omega_{ij},\,i,j=2,3$, solving the problem first raised by Aleksan {\em et al} \cite{AKL}. In the following we make the notation
\begin{eqnarray}
\tan \omega_{22}=t_{22},\quad\tan \omega_{23}=t_{23},\quad\tan \omega_{32}=t_{32},\quad\tan \omega_{33}=t_{33}\end{eqnarray}
and from the Eqs.(\ref{l5}),(\ref{l7}), and (\ref{l11})-(\ref{l12}) we get
\begin{eqnarray}
c_{13}^2&=&\frac{t_{23}\,t_{33}(t_{22}-t_{32})}{t_{22}\,t_{23}(t_{33}-t_{32})+t_{32}\,t_{33}(t_{22}-t_{23})}\nonumber\\
c_{23}^2&=&\frac{t_{32}\,t_{33}(t_{22}-t_{23})}{t_{22}\,t_{23}(t_{33}-t_{32})+t_{32}\,t_{33}(t_{22}-t_{23})}\label{f1}\\
c_{12}^2&=&\frac{N_1}{N_2}\nonumber
\end{eqnarray}
where
\begin{eqnarray}
N_1&={}&(t_{22}-t_{23})(t_{22}-t_{32})(t_{23}-t_{33})(t_{32}-t_{33})\nonumber\\
N_2&={}&t_{23}^2\,t_{32}^2+t_{22}^2\,t_{33}^2+t_{23}^2\,t_{32}^2(t_{22}^2+t_{33}^2)+t_{22}^2\,t_{33}^2(t_{23}^2+t_{32}^2)-\label{f2}\\
&{}\,& 2\,t_{22}\,t_{23}\,t_{32}\,t_{33}\left[1+ (t_{23}+t_{32})(t_{22}+t_{33})-t_{23}\,t_{32}-t_{22}\,t_{33}\right]\nonumber
\end{eqnarray}

Substituting the values for $c_{ij}$ from relations (\ref{f1}) and (\ref{f2}) in any equation (\ref{l5}),(\ref{l7}),(\ref{l11}), (\ref{l12}), or a combination of them, we get a formula for $\cos \delta$, that is too long to be written down here. Hence the necessary and sufficient conditions for the existence of a unitary matrix coming from the angles $\omega_{ij}$ are
\begin{eqnarray}0\le c_{12}^2\le1, \quad0\le c_{13}^2\le1, \quad0\le c_{23}^2\le1, \quad -1\le\cos\delta\le 1\end{eqnarray}

By using the numerical values
\begin{eqnarray}
t_{22}=\frac{1}{9}\sqrt{\frac{77}{3}},\,t_{23}=\sqrt{\frac{33}{7}},\,t_{32}=-\frac{\sqrt{231}}{13},\,t_{33}=\frac{3\sqrt{231}}{61}\end{eqnarray}
obtained from the matrix (\ref{unt}) we get by using the relations (\ref{f1}) and (\ref{f2})
\begin{eqnarray}c_{12}=\frac{1}{\sqrt{3}},\quad c_{13}=\frac{\sqrt{3}}{2},\quad c_{23}=\frac{\sqrt{6}}{4}\label{ang}\end{eqnarray}
showing that Eqs.(\ref{f1})-(\ref{f2}) uniquely define the parameters $c_{ij}$, in perfect accord with (\ref{del1}).

If in Eqs.(\ref{l5}),(\ref{l7}), and (\ref{l11})-(\ref{l12}) we substitute the angles as given by  (\ref{f1}) and (\ref{f2}) we get equations for $\cos\delta$
that lead to the solutions
\begin{eqnarray}\cos\delta&=&\frac{4}{5}\sqrt{\frac{3}{5}},\quad\cos\delta=-\frac{139}{116}\sqrt{\frac{3}{5}},\quad\cos\delta=\frac{9}{8}\sqrt{\frac{3}{5}},\quad\\
\cos\delta&=&-\frac{1}{4}\sqrt{\frac{3}{5}},\quad\cos\delta=-\frac{41}{32}\sqrt{\frac{3}{5}}\end{eqnarray}
In fact from each equation (\ref{l5})-(\ref{l7}) and  (\ref{l11})-(\ref{l12}) one gets two solutions for $\cos\delta$, and only one of them coincides with   that found in the first phenomenological model, see (\ref{del1}). Hence the problem of recovering a unitary matrix when we know four independent angles $\omega_{ij}$ is not unique, and the finite multiplicity is at least five. However this result does not contradict the general result stated in  Theorem 2, theorem  which gives  oneness then and only then when we use four independent moduli. To see what happens in the above case,  we recover the unitary matrix by using  $c_{ij}$ taken from the  relations (\ref{ang}), and the second value for $\cos\delta$
\[\cos\delta=-\frac{139}{116}\sqrt{\frac{3}{5}}\]
 The moduli matrix has the form
\begin{eqnarray}
V_1=\left(\begin{array}{ccc}
\frac{1}{3}&\frac{1}{2}&\frac{1}{6}\\*[2mm]
\frac{1}{4}&\frac{47}{1856}&\frac{1345}{1856}\\*[2mm]
 \frac{5}{12}&\frac{881}{1856}&\frac{605}{5568}\end{array}\right)\end{eqnarray}
By comparing the matrix $V_1$ with the original one, Eq.(\ref{toy3}), we see that the elements of the first row and column coincide, the others are different.  In order to obtain a unique solution we can use the relation (\ref{l8}), or any other equivalent to, which make use of one of the lengths of unitarity triangle sides.  Hence by using information coming only from triangles angles we have a finite multiplicity solution. The unicity is obtained then and only then when the information is supplemented by an independent  modulus, e.g. in the above case $V(2,2)$,  or a length of a unitarity triangle.  The phenomenological implications of the above results on the global fit methods for recovering a unitary matrix from moduli and angles will be treated elsewhere.

\section{Recovery of unitary matrices from experimental data}

If the data come from an exact numerical matrix   the problem to solve is quite simple: we have to test the stochasticity property, and, afterwords, the unitarity constraints, i.e. the condition, $-1 \le \cos\delta \le 1$, in the unitarity condition method, or the inequalities, $|R_{\alpha\beta,\gamma} -R_{\alpha\beta,\gamma}| \le 1 \le  R_{\alpha\beta,\gamma} +R_{\alpha\beta,\gamma}$, coming from two orthogonality relations in the case of the standard unitarity triangles approach. If the data pass anyone of the tests, one can easily reconstruct the unitary matrix Eq.(\ref{ckm}) from the data (\ref{pos}), as the numerical examples from the previous sections have shown. If the physical conditions are violated, there is no compatibility and the
discussion ends here. The real problem, from a physical point of view, is when the  data come from experiment, i.e.  are numbers affected by errors, and the problem is how we proceed  in this situation, because neither the double double stochasticity relations, nor the unitarity constraints are  exactly satisfied.

 There is the place where the gauge invariance subgroup $K$ of unitary matrices enters the game, and its implications are the following. We have to find all the {\em four} independent moduli groups and find all the possible forms for $\cos\delta$, and, respectively, for the lengths of the unitarity triangles. And we have to impose that the numerical values for them should be approximately equal. The usual case with the present data is that the numerical values obtained for moduli are such that  $\cos\delta^{(i)}\neq\cos\delta^{(j)} $, and/or  $R_{\alpha\beta,\gamma}^{(i)} \neq R_{\alpha\beta,\gamma}^{(j)},\,\, i\ne j$. Even more $\cos\delta^{(i)}$ could be outside the physical region, or the lengths of unitarity triangles are imaginary, or if they are real are not compatible with the existence of a triangle, as the numerical examples provided in the paper show. Hence, in contradistinction with the nowadays usage, see \cite{BaB}, we have to devise a  fitting model that should implement the fulfillment of the above theoretical constraints, and which should take into account the experimental data.

The method we expose here is discussed in more detail in \cite{Di1}. It is  a least squares method for checking the compatibility of the data with the theoretical models in both the approaches, and if the data pass the physical conditions imposed by unitarity, from the fits one gets values for the parameters entering the theoretical model that by assumption  gives a reliable description of the physical reality that is  investigated by experiment.

It follows that  in both the approaches the $\chi^2$-function must contain two separate terms, the first have to impose the fulfillment of the unitarity constraints by  the free parameters entering the physical  model, and their best determination from   the data, and the second one should depend on physical quantities that are measured in (different) experiments. Thus our proposal for the first terms is
 
\begin{eqnarray}
\chi^2_1=\sum_{i < j}(\cos\delta^{(i)} -\cos\delta^{(j)})^2+\sum_{j=u,c,t}\left(
\sum_{i=d,s,b}V_{ji}^2-1\right)^2\nonumber\\
+\sum_{j=d,s,b}\left(
\sum_{i=u,c,t}V_{ij}^2-1\right)^2,\,\,\,\,-1\le\cos\delta^{(i)}\le 1\label{chi} 
\end{eqnarray}
for the unitarity condition method, and, respectively,

\begin{eqnarray}
\chi^2_2&=&\sum_{\substack{\alpha\beta,\gamma \\ i <j}} (R_{\alpha\beta,\gamma}^{(i)} -R_{\alpha\beta,\gamma}^{(j)})^2+
 \sum_{j=u,c,t}\left(
\sum_{i=d,s,b}V_{ji}^2-1\right)^2
+\sum_{j=d,s,b}\left(
\sum_{i=u,c,t}V_{ij}^2-1\right)^2, \nonumber \\*[2mm]
&{}&R_{\alpha\beta,\gamma}^{(i)}\ge 0,\quad |R_{\alpha\beta,\gamma}^{(i)} -R_{\alpha\beta,\gamma} ^{(i)}|\le 1\le R_{\alpha\beta,\gamma}^{(i)}+R_{\alpha\beta,\gamma}^{(i)} \label{chi-3} 
\end{eqnarray}
for the unitarity triangles method. Both the $\chi^2_{1,2}$ formulas test the double stochasticity property and the unitarity; from the point of view of numerical computation the unitarity property is the most difficult to satisfy.

Concerning the second component of $\chi^2$-test it is of the form
\begin{eqnarray}
\chi^2_3=\sum_{i=1}\left(\frac{d_i-\widetilde{d_i}}{\sigma_i}\right)^2\label{chi1}\end{eqnarray}
where $d_i$ are theoretical functions depending on the theoretical parameters $s_{ij}$ and $\delta$, or on $V_{kl}$, or/and the angles $\phi_i,\psi_i$  which are given by the phenomenological model one works with, while  $\widetilde{d_i}$ are the measured experimental data for $d_i$, and $\sigma$ is the vector of errors associated to $\widetilde{d}_{i}$. The formulas 
\[\chi^2_u=\chi^2_1+\chi^2_3\] 
 and, respectively,
\[\chi^2_t=\chi^2_2+\chi^2_3\]
are our proposals for $\chi^2$-tools necessary in  analyzing the experimental data. 

A remark is the following: as we have seen the second phenomenological model does not provide a formula for $\cos\delta$. However any global fit done by using either $\chi^2_u$ or $\chi^2_t$ gives values for all the moduli. Hence for the reconstruction of  a unitary matrix in the second  phenomenological model we have to use  formulas such as (\ref{l8}).  A true global fit will be that  which  will use  all the experimental data by merging the above  two  phenomenological models.  

The convexity property of the double stochastic matrices allows us to devise a method for doing statistics on unitary matrices that is still an open problem in the physical literature. Let suppose that by doing a fit with the above methods we got $n$ moduli matrices that are consistent with $n$ (approximate) unitary matrices,    $U_1,\,U_2\cdots,U_n$. The convexity property together with the embedding (\ref{ds1}) tell us that the matrix
\[M^2=\sum_{i=1}^{n} x_i\,|U_i|^2, \quad \sum_{i=1}^{n} x_i=1,\quad0\le x_i\le 1, \,\, i=1,\cdots,n\]
is double stochastic, where we use the Hadamard product, so the above relation is understood as working entry wise. Then the  correct formulas for the mean value $< M >$, and the error matrix $\sigma_M$ are given by
\begin{eqnarray}
< M >&=&\sqrt{(\sum_{i=1}^n\,|U_i|^2)/n}\nonumber\\
\sigma_M&=& \sqrt{(\sum_{i=1}^n\,|U_i|^4)/n-<M>^4}
\end{eqnarray}
If the entries of the mean value matrix, $< M >$, obtained in this way are not too far from the entries coming from a unitary matrix, one can reconstruct from  $< M >$ an (approximate) unitary matrix by using the technique developed in the paper.

\section{Conclusions}
Our main reason for studying the separation criteria between the double stochastic matrices and the unistochastic ones was their importance in high energy physics, where the nowadays algorithms for reconstruction of unitary matrices from experimental data are not yet reliable in our opinion. Fortunately for the $3\times 3$ matrices, that seems to be the physical choice in the electroweak interaction, there are explicit formulas for the independent parameters entering a unitary matrix in terms of four independent elements of a double stochastic one. That allows us to check the unitarity properties of exact double stochastic matrices, and an easy reconstruction of the unitary one from the entries of the double stochastic matrix when the compatibility conditions are fulfilled. These formulas can be used to define $\chi^2$-functions for  checking  the compatibility between the experimental data and the unitarity property of the CKM matrix, and to recover a unitary matrix from error affected data. More important, starting from the convexity of the Birkhoff's polytope, we found a method for doing statistics on the (moduli of) unitary matrices.

We have  also shown that, because the unitarity triangles method \cite{BaB} and the unitarity condition method \cite{D5}, being  both consequences of the unitarity property, are completely equivalent when and only when they are formulated in terms of  four independent moduli. In the same time we have shown that  the unitarity triangles method has to make effective use of the double stochasticity relations in order to obtain reliable results. Writing the unitarity triangles method in complex form we have obtained  formulas for  the four independent phases entering a unitary matrix, these phases being the angles of the unitarity triangles. This   opens the possibility to treat coherently all the experimental data available on moduli and angles by merging the above phenomenological models into a true global one, the aim in view being a precise   determination of the phase $\delta$ that is the key parameter in understanding the CP-violation.


\begin{thebibliography}{99}
\bibitem{Bir} D. Birkhoff, {\em Tres observaciones sobre el algebra lineal } Univ.Nac.Tucum\'an Rev, {\bf A5} (1946) 147-151
\bibitem{MO} A. W. Marshall and I. Olkin, {\em Inequalities: Theory of Majorization and Its Applications}, (Academic Press, New York, 1979), Chapter 2
\bibitem{YP} Y.-H. Au-Yeung and Y.-T. Poon,  $3\times 3$ Orthostochastic Matrices  and the Convexity of Generalized Numerical ranges, Lin.Alg.Appl. {\bf 27} (1979) 69-79
\bibitem{N} H. Nakazato, Set of $3\times 3$ Orthostochastic Matrices, Nihonkai Math.J. {\bf 7} (1996) 83-100 
\bibitem{MN} G. Mennessier and J. Nuyts, Some unitary bounds for finite matrices, J.Math.Phys. {\bf 15} (1974) 1525-1537
\bibitem{A} G. Auberson, On the reconstruction of a unitary matrix from its moduli. Existence of continuous ambiguities, Phys.Lett. {\bf B 216} (1989) 167-171
\bibitem{L1} L. Lavoura, On the reconstruction of the four-generation CKM matrix from the moduli of its matrix elements, Phys.Lett. {\bf B 223} (1989) 97-102
\bibitem{L2} L. Lavoura, Parametrization of the four-generation quark mixing by the moduli of its matrix elements, Phys.Rev. {\bf D 40} (1989) 2440-2448
\bibitem{AMM} G. Auberson, A. Martin, and G. Mennessier, On the reconstruction of a unitary matrix from its moduli, Commun.Math.Phys. {\bf 140} (1991) 417-437
\bibitem{YC} Y.-H. Au-Yeung and  C.-M. Cheng, {Permutation matrices whose convex combinations are orthostocastic}, Lin.Alg.Appl. {\bf 150} (1991) 243-253
\bibitem{D1} P. Di\c t\u a, Parametrization of unitary matrices by moduli of their elements, Commun.Math.Phys. {\bf 159} (1994) 581-591
\bibitem{BEKTZ} I. Bengtsson, A. Ericsson, M Ku\'s, W. Tadej, and K. \.Zyczkowski, Birkhoff's polytope and unistochastic matrices, N=3 and N = 4,
 Commun.Math.Phys. {\bf 259} (2005) 307-324

\bibitem{P} M. Petrescu, {\em Existence of continuous families of complex Hadamard matrices of certain prime dimensions and related results}, UCLA thesis, (1997), Los Angeles
\bibitem{UH} U. Haagerup, Orthogonal maximal abelian $*$-subalgebra of the
  $n\times n$ matrices and cyclic $n$-roots,  in {\em Operator algebras and
    quantum field theory} Rome, 296-322, Internat.  Press, Cambridge,
  MA, 1997
\bibitem{D2} P. Di\c t\u a, Some results on the parameterization of complex Hadamard matrices, J.Phys.A: Math.Gen. {\bf 37} (2005) 5355-5374
\bibitem{Ni} R. Nicoara, A finiteness result for commuting squares of matrix algebras, {\em preprint} math.OA/0404301
\bibitem{TZ} W. Tadej and K. \.{K}yczkowski, A concise guide to complex Hadamard matrices, {\em preprint} quant-ph/0512154

\bibitem{Ed} S. Eidelman   {\em et al.}, Review of the particle physics,  Phys.Lett. {B \bf 592} (2004) 1-1109
\bibitem{KM} M. Kobayashi and T. Maskawa, CP-violation in the renormalizable theory of weak interaction, Prog.Theor.Phys. {\bf 49} (1973) 652-657
\bibitem{BaB} {\em The BaBar Physics Book}, P.H. Harrison and H.R. Quinn (eds),
SLAC-R-504, Ch. 14 (1998)


\bibitem{BLO} A. J. Buras, M. E. Lautenbacher and G. Ostermaier, Waiting for the top quark mass, $K^+\rightarrow p^+\nu\overline{\nu}$, $B_s^0-\overline{B}_s^0$ mixing and $CP$ asymmetries in $B$ decays, Phys.Rev. {\bf D 50} (1994) 3433-3446
\bibitem{CL} M. Ciuchini et al., 2000 CKM-triangle analysis. A critical review with updated experimental inputs and theoretical parameters, JHEP {\bf 0701} (2001) 013

\bibitem{HLLL} A. H\"ocker, H. Lacker, S. Laplace, and F. R. Le Diberder, A new approach to a global fit of the CKM matrix,  Eur.Phys.J. {\bf C21} (2001) 225
\bibitem{BPS} A. Buras, F Prodi, and A Stocchi, he CKM matrix and the unitarity triangle: another look, JHEP {\bf 01} (2003) 029
\bibitem{JC} J. Charles  {\em et al.} (The CKM Fitter Group), CP violation and the CKM matrix: assessing the impact of the asymmetric B factories, Eur.Phys.J. {\bf C 41} (2005) 1-131; hep-ph/0406184
\bibitem{Bo} M. Bona et al., (UTfit collaboration), The 2004 UTfit collaboration report on the status of the unitarity triangle in the standard model,  JHEP {\bf 07} (2005) 029; hep-ph/0501199
\bibitem{St} A. Stocchi, Current status of the CKM matrix and the CP violation,  {\em preprint} hep-ph/0405038
\bibitem{RF} R. Fleischer, Flavour physics and CP violation, {\em Lectures given at the 2003 European School of High-Energy Physics, Tsakhkador, Armenia, 24 August-6 September 2003}, {\em preprint},  hep-ph/0405091
\bibitem{B} A. Buras, Flavour Physics and CP violation, {\em Lectures given at the European CERN School, Saint Feliu Guixols, June 2004},  hep-ph/0505175
\bibitem{Fl} R. Fleischer, New physics in $B$ and $K$ decays, {\em  Invited lectures given at Lake Louise Winter Institute: ``Fundamental Interactions'', Chateau Lake Louise, Alberta, Canada, 20-26 February 2005 }, {\em preprint}, {\em preprint},  hep-ph/0505018




\bibitem{M} F.D. Murnagham, {\em The Unitary and Rotation Groups} (Washington, DC: Sparta Books)   (1962)
 \bibitem{D3} P. Di\c t\u a, Parametrization of unitary matrices, J.Phys.A: Math.Gen. {\bf 15} (1982) 3465-3473
 \bibitem{CJ} C. Jarlskog, Commutator of the quark mass matrices in the standard electroweak model and a measure of maximal CP non conservation, Phys.Rev.Lett. {\bf 55} (1985) 1039-1042
 \bibitem{BD} J.D. Bjorken and I. Dunietz, Rephasing invariant parameterizations of generalized Kobayashi-Maskawa matrices, Phys.Rev. {\bf D 36} (1987) 2109-2118
 \bibitem{NP} J.F. Nieves and P.B. Pal, Minimal rephasing-invariant CP-violating parameters with Dirac and Majorana fermions, Phys.Rev. {\bf D36} (1987) 315-317
\bibitem{Br} G. C. Branco and L. Lavoura,  Rephasing-invariant parametrization of the quark mixing matrix, Phys.Lett.  {\bf B 208}  (1988) 123-127
\bibitem{CK} L.L. Chau and W Y Keung, Comments on the parametrization of the Kobayashi-Maskawa matrix, Phys.Rev.Lett. {\bf 53} (1984) 1802-1805
\bibitem{D4} P. Di\c t\u a, Factorization of unitary matrices, J.Phys.A: Math.Gen. {\bf 36} (2003) 2781-2789
\bibitem{J1} C. Jarlskog, A recursive parameterization of unitary matrices, J.Math.Phys. {\bf 46} (2005) 103508; Recursive parameterization and invariant phases of unitary matrices, {\em preprint} math-ph/0510034
\bibitem{CM} S.Chaturvedi and N.Mukunda, Parameterizing the mixing matrix; a unified approach, Int.J.Mod.Phys. {\bf A 16} (2001) 1481-1490
\bibitem{MS} M.Mathur and D.Sen, Coherent states for $SU(3)$, J.Math.Phys. {\bf 42} (2001) 4181-4186

\bibitem{HF} http://www.stanford.edu/xorg/hfag/triangle/

\bibitem{D5} P. Di\c t\u a, Another method for a global fit of the Cabibbo-Kobayashi-Maskawa matrix, Rom.J.Phys. {\bf 50} (2005) 279-287
\bibitem{AKL} R. Aleksan, B. Kayser and D. London,  Determining the quark mixing matrix and $CP$-violating asymmetries, Phys.Rev.Lett. {\bf 73} (1994) 18-20
\bibitem{W} L. Wolfenstein, Parametrization of the Kobayashi-Maskawa matrix, Phys.Rev.Lett. {\bf 51} (1983) 1945-1947


\bibitem{Di1} P. Di\c t\u a, Modern Phys.Lett., Global fits to the Cabibbo-Kobayashi-Maskawa matrix: unitarity condition method versus standard unitarity triangles approach,     { \bf A  20} (2005) 1709-1721




\end{thebibliography}
\end{document}